# Chronic, cortex-wide imaging of specific cell populations during behavior

Joao Couto[1,2,10], Simon Musall[1,3,4,10], Xiaonan R Sun[1,5], Anup Khanal[1,2], Steven Gluf[1],
Shreya Saxena[6,7,8,9], Ian Kinsella[6,7,8,9], Taiga Abe[6,7,8,9], John P. Cunningham[6,7,8,9],
Liam Paninski[6,7,8,9], Anne K Churchland[1,2,*]

[1] Cold Spring Harbor Laboratory, Neuroscience, Cold Spring Harbor, NY, USA
[2] Department of Neurobiology, University of California, Los Angeles, CA, USA
[3] Institute of Biological Information Processing (IBI-3), Forschungszentrum Jülich, Jülich, Germany
[4] Department of Neurophysiology, Institute of Biology 2, RWTH Aachen University, Aachen, Germany
[5] Department of Neurosurgery, Zucker School of Medicine, Hofstra University, Hempstead, NY, USA
[6] Mortimer B. Zuckerman Mind Brain Behavior Institute, Columbia University, New York, NY, USA
[7] Department of Statistics, Columbia University, New York, NY, USA
[8] Center for Theoretical Neuroscience, Columbia University, New York, NY, USA
[9] Grossman Center for the Statistics of Mind, Columbia University, New York, NY, USA
[10] Equal contribution
[*] Correspondence: achurchland@mednet.ucla.edu

**Abstract:** Measurements of neuronal activity across brain areas are important for understanding the neural correlates of cognitive and motor processes like attention, decision-making, and action selection. However, techniques that allow cellular resolution measurements are expensive and require a high degree of technical expertise, which limits their broad use. Widefield imaging of genetically encoded indicators is a high throughput, cost effective, and flexible approach to measure activity of specific cell populations with high temporal resolution and a cortex-wide field of view. Here we outline our protocol for assembling a widefield setup, a surgical preparation to image through the intact skull, and imaging neural activity chronically in behaving, transgenic mice that express a calcium indicator in specific subpopulations of cortical neurons. Further, we highlight a processing pipeline that leverages novel, cloud-based methods to analyze large-scale imaging datasets. The protocol targets labs that are seeking to build macroscopes, optimize surgical procedures for long-term chronic imaging, and/or analyze cortex-wide neuronal recordings.

## Introduction

Simultaneous recordings of activity across brain areas have recently enabled unprecedented insights into the coordination of neuronal activity during both spontaneous[1–3] and cognitive behaviors[4–8]. One of the first methods used to unravel the spatial organization of large-scale cortical activity was widefield imaging[9,10]. Originally, hemodynamic signals were often used as a proxy for neuronal activity, and the large field of view and non-invasive nature of widefield imaging made it ideal for mapping cortical responses to sensory stimulation. However, hemodynamic signals have limited spatiotemporal resolution and specificity[11–14]. An alternative is using fluorescent activity indicators, such as calcium- or voltage-indicators, which can be genetically encoded, have high signal to noise ratio and provide direct insights into the activity of specific cell-types[5,15–17]. The widespread availability of transgenic mouse lines combined with advancements in recording techniques allow imaging the entire dorsal cortex with high spatial and temporal resolution.

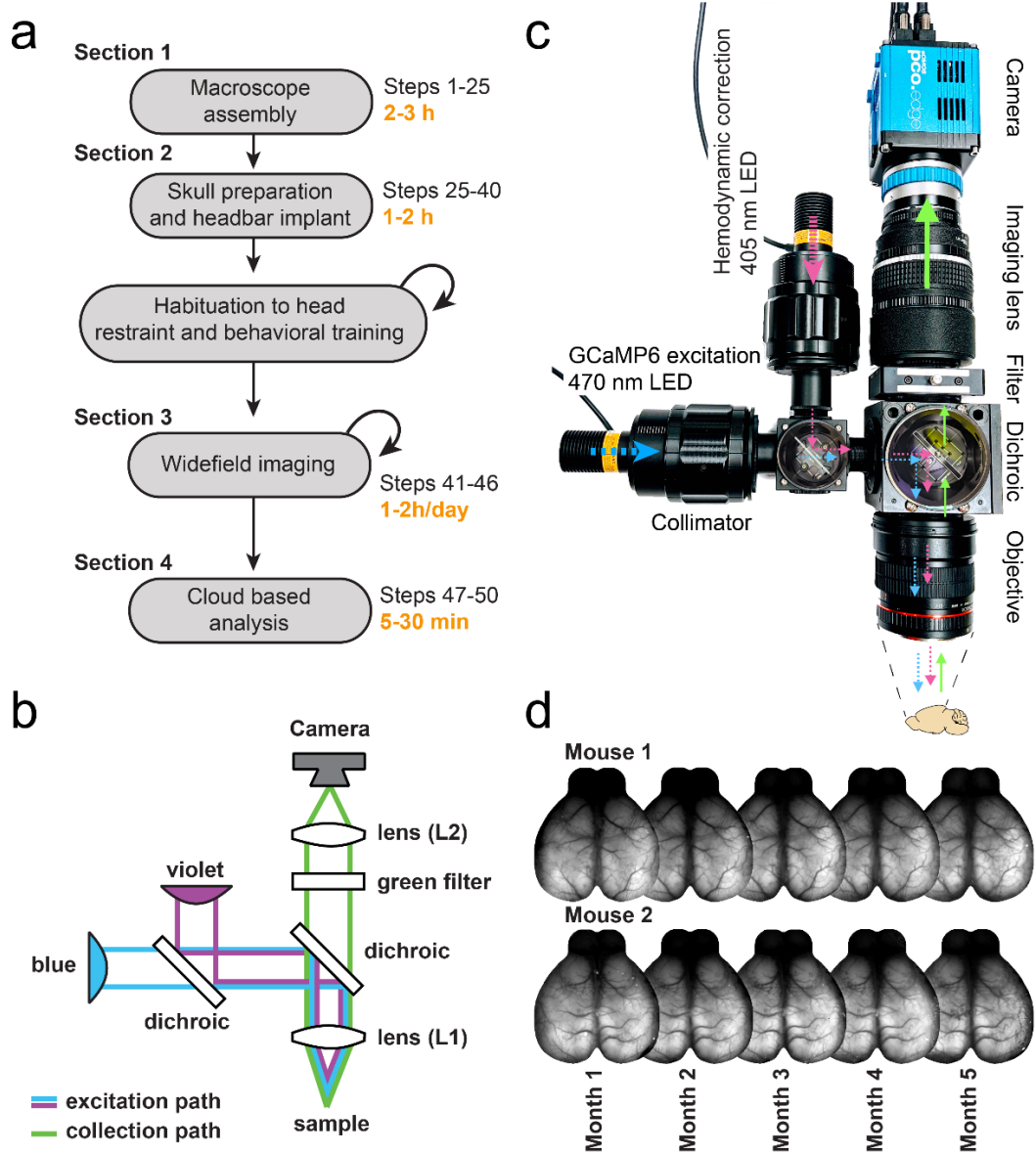

*Figure 1 **Overview of the procedures described in this protocol.** a) Main stages of a widefield experiment. Description of the animal preparation in steps 25-40; imaging in steps 41-46, and data preprocessing in steps 47-50. Animal habituation depends on the behavioral assay and is not discussed here. Data analysis beyond preprocessing depends on the scientific question; see the anticipated results section for examples. b) Diagram of the macroscope in tandem-lens configuration. c) Overview of the macroscope described in steps 1-25 with light path diagram. Colors indicate the wavelength of the light in the corresponding path (violet: 405nm; blue: 470nm). d) Implant stability over 5 months. Cyanoacrylate maintains the skull semi-transparent chronically, enabling long term monitoring of neuronal activity.*

The ability to record neural activity from a large number of cortical areas simultaneously with genetically encoded indicators enabled exciting possibilities beyond sensory mapping that have just recently come to the forefront. In particular, recordings from the entire dorsal cortex profoundly affected how we consider trial-to-trial variability and neural engagement during cognitive behaviors[3–7,18]. Alongside developments of neural activity indicators, new surgical preparations enable acquiring neural signals through the intact skull[19,20], thereby reducing the potential for brain trauma, inflammation and tissue regrowth. This makes widefield imaging ideally suited for longitudinal studies and high-throughput assays. However, a major bottleneck is that widefield imaging produces massive datasets that are not trivial to preprocess, analyze, store and share with the community. Recent advances in computational methods, such as techniques for dimensionality reduction and statistical modeling, provide new opportunities for compressing, denoising and exploring widefield data[21–23].

Here, we provide a protocol for assembling a macroscope for widefield imaging along with a step-by-step guide for the surgical preparation, data acquisition and analysis that in combination, allow repeated imaging of most of the dorsal cortex of transgenic animals expressing calcium indicators. The protocol consists of four sections (Fig 1a). First, we provide instructions for how to build and calibrate a widefield imaging macroscope. Second, we describe our surgical procedures for chronic imaging through the intact mouse skull. Third, we discuss how to acquire neural data and include data acquisition software to record from awake mice. In the last section, we provide instructions for deploying state-of-the-art computational algorithms for pre-processing and analysis of widefield data that take advantage of an cloud platform and focus on reproducibility and scalability. Taken together, this protocol aims to circumvent the main barriers to implementing longitudinal imaging of cortex-wide neuronal activity by introducing a novel, overarching platform that allows any investigator to build a widefield setup, measure neural activity, and analyze signals on the cloud, thus readily generating new observations about widespread cortical activity.

**Development and applications of the protocol**

The protocol emerged from the need to simultaneously and repeatedly record neural activity from distant cortical areas in head-restrained mice during perceptual tasks. This demanded combining a highly sensitive macroscope with a surgical preparation for optical access to the entire dorsal cortex. Further, it prompted the development of software for synchronizing imaging data with behavior, and the implementation of a scalable data analysis pipeline.

**Section 1: Macroscope assembly.** We based our macroscope on the design from Ratzlaff et al. that allows efficient light collection and excitation using a tandem-lens configuration[24]. The design uses commercially available parts, and is customizable, easy to assemble, and relatively cost-effective (under US$ 10k, excluding the camera), which places it within reach of laboratories without previous imaging expertise or with limited resources.

A widefield macroscope consists of two main components: the excitation path (blue and violet in Fig. 1b) that delivers light to the cortical surface to excite the fluorescent indicator; and the collection path (green in Fig. 1b) that collects light emitted by the fluorescent indicator and terminates at the camera sensor. In our design, the excitation path is optimized for imaging green fluorophores (e.g. GCaMP6) and uses a set of dichroic mirrors to direct light from the two LEDs to the objective. An alternative is to provide illumination externally from the side of the objective[25–27]. Side illumination requires fewer components and allows the flexibility to use different wavelengths (e.g. when green excitation light[17] or multiple wavelengths are required[27]); however its implementation can be challenging when using visual stimulation assays where light shielding is important.

The sensitivity of the collection path is an important factor that determines if weaker signals can be recorded and how much light power needs to be delivered to the cortex. Light collection in our design uses two single-lens reflex camera (SLR) lenses that are facing each other in an inverted tandem-lens configuration to form a low-magnification collection path[24]. This provides long working distance (>40 mm – dictated by the flange focal distance of the objective lens), narrow depth of field that maximizes signal collection from cortex, and high sensitivity (numerical aperture of ~0.36 versus ~0.05 for low magnification microscope objectives). Alternative designs, using a single camera lens[26] or a microscope lens[28] can also be used to reduce cost or setup size but often results in shorter working distances, lower sensitivity and wider depth of field.

For the tandem-lens configuration, the objective lens is inverted (i.e. the lens' camera mount faces the brain surface) to project near-collimated light to a second lens that focuses the light on the camera sensor. The ratio of the focal lengths of two lenses, L1 and L2 in Fig. 1b, determines the magnification. We chose L1 f=105 mm for the imaging (top) lens and L2 f=85 mm for the objective (bottom lens) which results in a magnification of 1.24 (L1f/L2f). To increase or decrease the magnification factor, one can use lenses with different focal lengths. It is important to consider that the effective field of view depends on the magnification factor and the size of the camera sensor. We use a camera with a 16.6 x 14 mm sensor, which results in a field of view of 13.4 x 11.3 mm with a 1.24 magnification factor. Because both lenses include zoom optics, it is possible to adjust the field of view, without replacing the lenses, between 17.1 x 14.4 mm (26.7 μm/pixel) and 12.5 x 10.5 mm (19.4 μm/pixel) which is ideal for imaging the whole cortex of adult mice. To obtain high signal senstivity and a large field of view, we use a camera (PCO Edge 5.5) with a scientific CMOS sensor that provides high-quantum efficiency (~ 60% for green light), large sensor (16.6 by 14 mm), low noise, large dynamic range (16 bit), fast pixel scan rate (286 MHz) and high frame rates (100Hz at full resolution). The shutter mode of the camera is also an important consideration when choosing the camera because acquiring signals from moving targets with a rolling shutter can produce artifacts. We further discuss how to acquire in rolling shutter mode in the third section. It is also possible to use cameras with lower sensitivity depending on the experimental constrains (brightness of the fluorescent indicator and required sensitivity). In earlier versions of the setup, we used the MV-D1024E-160-CL-12 camera from Photonfocus with satisfactory results; recent studies even demonstrated cortex-wide imaging using a lower-cost Raspberry Pi Camera module[26,29].

To maximize the amount of collected light we use a 60mm cage system (hollow cube where optical components – lenses, mirrors and filters – are mounted) in the collection path. The ratio between the focal length and the diameter of the entrance pupil (effective aperture) is the f-stop (or f-number) and is used in photography as a measure of the aperture in different lenses. Our objective lens (L1) has an f-stop of 1.4 and a focal length of 85mm. The entrance pupil is therefore 85mm / 1.4 = 60.7 mm. Using a 60mm, and not a 30mm, cage system is an important design consideration because it allows more light to be collected by the imaging lens (L2). Another consideration is that the light exiting the L1 lens is slightly divergent and therefore the entrance pupil of L2 should be larger than the exit pupil of L1[24]. In our design, the exit pupil of L1 has a 50mm aperture (the size of optical mounts in our cage system) and the L2 lens has entrance pupil of 105 mm / 2 = 52.5 mm.

The near-collimated light path between the objective (L1) and the imaging (L2) lenses can accommodate optics specific to fluorescence imaging (Fig. 1b,c). We use a dichroic mirror to reflect blue (470 nm) and violet light (405 nm) on the sample while collecting the emitted, longer wavelength (>495 nm) light through a green band-pass filter (525 ± 25 nm) (Fig. 1b, c). These optics are optimized for imaging green fluorophores (e.g. GCaMP6) but can easily be exchanged to record with other activity indicators, e.g. red-shifted indicators.

The acquired fluorescence is a combination of neuronal signals from the activity indicator but also non-neuronal signals from other sources, such as flavoprotein- and hemodynamic-related fluorescence changes [27,30]. To isolate neuronal signals, we alternate the excitation light from frame to frame between 470nm (blue light) and 405nm (violet light) [4,5,30]. While blue light excites GCaMP to emit green light depending on the binding of calcium, violet light is at the isosbestic point for GCaMP, which means that the emitted fluorescence is largely independent of the calcium-binding. Since fluorescence at both wavelengths contain non-neural signals but only blue-excited fluorescence additionally contains neural signals, we can rescale violet-excited fluorescence to match blue-excited fluorescence and then subtract it to remove non-neural signals[4] (Fig. 3). This effectively removes intrinsic signal contributions and has been validated in single neurons[30] and GFP-expressing control mice[4]. Nonetheless, a consensus on the optimal strategy for hemodynamic correction has not yet been reached. Efforts to further quantify different components of the hemodynamic response, its relation to GCaMP fluorescence and the development of indicators that are less prone to contamination are important to find novel ways of accurately estimating calcium fluorescence[27,31]. An alternative approach, which may reduce photobleaching, is to use green instead of violet illumination[17]. A second alternative is to use combinations of blue, green and red illumination to individually quantify hemodynamic signals, such as fluorescence changes due to oxygenation of hemoglobin and local blood volume[27]. The approaches discussed above require alternating the excitation light from frame to frame between different wavelengths; a related approach is to use continuous illumination with two wavelengths and separate the emitted light using an RGB camera[29], or building a spatial model of the hemodynamic response using a separate cohort of non-GCaMP mice[31].

In some cases, the hemodynamic correction can also be omitted[32–34]. For example when imaging from densely labeled neuronal populations, using bright indicators like GcaMP6s, the neural signal can be strong enough to not be confounded by non-neural signals[32,33,35]. However, since intrinsic signal contamination strongly increases the more excitation and emission photons travel through brain tissue[27], the correction is particularly important when imaging from deeper cortical layers even when using bright neural indicators.

**Section 2: Animal preparation.** Imaging cortical activity requires a preparation that ensures optical access to the cortex for the duration of the experiment. A specific procedure, often referred to as 'skull clearing'[19], allows to maintain optical clarity similar to that observed when covering the skull with saline over months without chemically altering the bone structure[20]. We favor preparing the skull with cyanoacrylate[19] over invasive procedures[36,37] because of the low risk of tissue damage, straight-forward surgical procedure and long-term implant stability. The cyanocrylate we use was selected for three reasons: it prevents the formation of opaque spots that are due to locking in small air bubbles in the bone, it has a medium viscosity that allows even distribution across the skull, is very clear after curing, and remains stable to allow repeated imaging for at least 5 months (Fig. 1d).

When imaging from specific cell populations, brightness and expression of the indicator in the neuronal population of interest is critical. For imaging exctitatory neurons in cortex, the transgenic tetO-GCaMP6s/CaMK2a-tTA mouse line[17] provides very high signal strength and even expression across cortical areas[38]. In contrast, some GCaMP6-expressing mouse lines, such as the Ai93D and Ai95D lines, have weaker fluorescence signal and can exhibit epileptiform activity patterns[39].

**Section 3: Data acquisition.** Several considerations are important regarding data acquisition when imaging repeatedly from transgenic mice. First, most high-speed CMOS cameras operate in 'rolling shutter' mode. However, this can cause artifacts when neural dynamics change quickly because not all lines are exposing at the same time. To prevent such artifacts, we restrict the excitation light to times when all lines are exposing. To do this, the camera common exposure trigger is connected to a fast microcontroller (Teensy) that triggers the LEDs in an alternating fashion and only during the exposure time common to all lines (Fig.2a and b). It is therefore important to use a light shield to prevent contamination of the imaging signals with ambient light because some lines are partially exposing during the time that the excitation light is turned OFF. For single wavelength illumination, the camera common exposure trigger can be connected to the LED directly. Note that this is only critical for cameras with 'rolling shutter' acquisition but does not apply to cameras that operate in 'global shutter' mode.

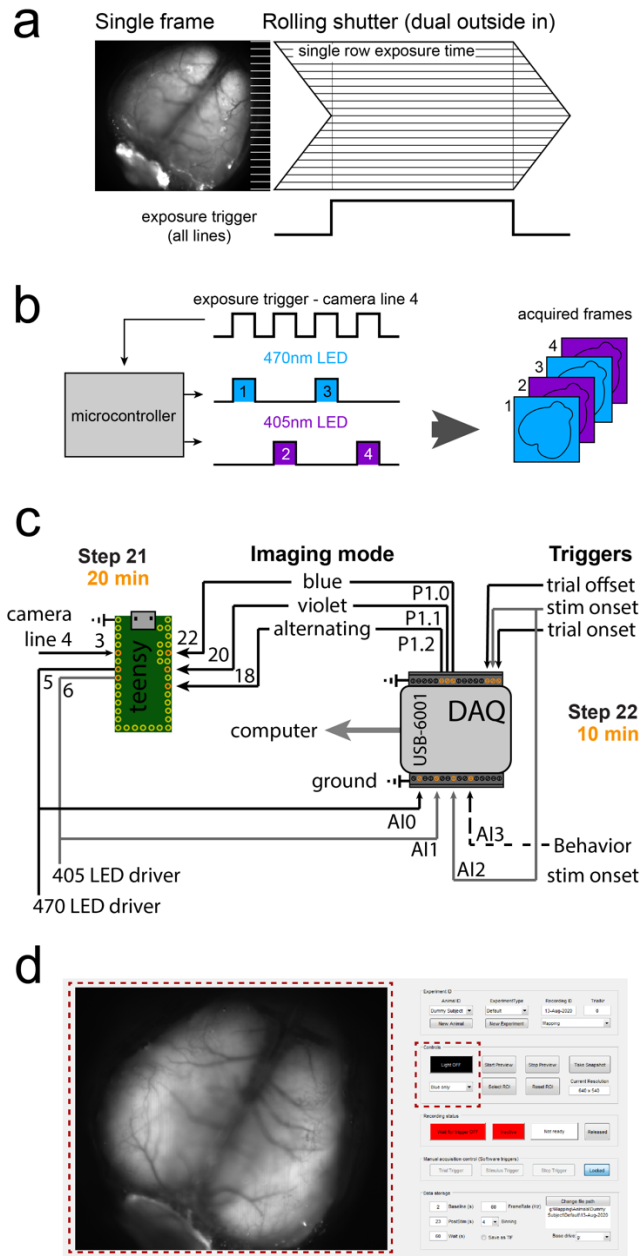

*Figure 2 **Acquisition in rolling shutter, dual color excitation mode and synchronization** a) Illustration of single frame acquisition with a rolling shutter camera in "dual outside-in" mode. Common exposure time of all lines is restricted to part of the frame duration. The dorsal cortex is in focus as in step 44 for imaging. b) Excitation light is restricted to the period of common exposure of all lines and alternated between 470nm and 405nm. The microcontroller alternates light according to the camera exposure input. c) Connection diagram for the microcontroller circuit and integration with the data acquisition system for synchronizing with a behavioral setup. Dashed line is optional. Gray lines are to disambiguate overlap. d) Graphical user interface of the acquisition software. Red, dashed squares highlight the preview (left) and LED control (right) panels. A detailed description of the software is in https://github.com/musall/WidefieldImager.*

Second, imaging for long durations can induce photobleaching, the photochemical degradation of the fluorophore, which reduces the signal amplitude over time. In addition to optimizing the light sensitivity so that less power is required, and triggering LED illumination with the camera, we devised a trial-based acquisition system in which the LEDs are extinguished during the period between trials. Together, this strategy greatly reduces the net power delivered to cortex by restricting light exposure periods and thus enabling chronic imaging across months without signal decay. We usually use 10 to 25 mW of alternating blue/violet light when imaging at 30 Hz. The required light power depends on the brightness, expression and labeling density of the indicator and therefore varies from mouse to mouse. We observed that light powers of 50 mW and higher can induce photobleaching within 5-10 minutes of imaging (Fig S1). Photobleaching is reversible due to the regeneration of indicator protein over time but bleaching effects can also accumulate when imaging daily (Fig S1). If bleaching occurs within a session, waiting several days between imaging sessions can avoid such accumulation.

Third, recording behavioral events and stimulus triggers alongside imaging data is critical for the accurate interpretation of neural activity. We use a data acquisition board (DAQ) to acquire the exposure for each frame and the LED triggers, and to monitor behavior variables (e.g. from a treadmill or from sensors in the behavioral setup). An off-line algorithm then asserts proper alignment of behavioral and imaging data. The data acquisition board also controls the microcontroller (Fig. 2c) that controls the excitation light for multiple wavelength illumination and is used to recover the wavelength that was used for each frame. An alternative approach is to use a camera that supports 'general purpose input-output' (GPIO) to record the triggers directly in the camera frames or trigger the camera to acquire a pre-determined number of frames when the trial structure is known *a priori*.

Lastly, widefield imaging produces very large datasets (n the order of ~150GB/h when imaging continuously at 60Hz – depending on the resolution) which may be prohibitively expensive to store. However, since behavioral tasks often have periods during which data are not critical, we established an approach to discard the inter-trial periods even when the subject itself initiates the trials in the assay. This is a challenge because in self-initiated tasks, the baseline needs to be captured before the trial is initiated. To solve this problem we implemented a sliding window recording method that allows discarding the totality of the inter-trial interval with the exception of the baseline period and effectively relaxes the storage requirements (effectively cutting the length of the dataset by a factor of ~3 depending on the behavioral task).

To tackle all the aforementioned considerations, we developed a custom MATLAB-based acquisition software available in https://github.com/musall/WidefieldImager. The software aims to provide user-friendly control of the camera acquisition through a graphical user interface and to integrate with different behavior assays or stimulation paradigms (Fig. 2d). We also recently developed acquisition software written in Python with similar functionality (https://bitbucket.org/jpcouto/labcams). Furthermore, we provide scripts to evaluate the amount of bleaching that can be useful for repeated measurements from the same mouse (checkSessionBleaching.m).

**Section 4: Data analysis**. A key challenge for implementing the technique is the data analysis. Chronic widefield imaging of transgenic mice is a high throughput technique that produces massive datasets. Specific analysis algorithms may depend on the scientific question or experiment at hand. Here we dicuss some common processing steps to extract neuronal activity from the acquired frames. The first step is to correct for artifacts resulting from motion of the skull in relation to the objective. These can be corrected by aligning each frame to a time-averaged reference image. We obtained better results with rigid body registration algorithms that consider both translation and rotation artifacts. If there are mutliple channels, it is essential that motion is corrected for all channels because movement artifacts can be amplified when subtracting activity across channels. The next step is to compute relative fluorescence changes (ΔF/F) for each pixel. This is done by subtracting and dividing the value at each frame by the baseline fluorescence. The baseline can be computed by taking the average of a number of frames before trial onset or by taking the average of all frames in the session if the former is not possible.

The third step is to denoise and compress the dataset to improve the signal-to-noise ratio and separate the dataset into temporal and spatial components of reduced size. This simplifies subsequent analysis (e.g. linear decoding) because that can be performed on temporal components instead of individual pixels. Denoising aims to isolate the signal to boost signal-to-noise, whereas compression reduces the size of the dataset and is critical for the computational efficiency of downstream analyses. Here, we discuss two related methods for denoising and compression. The standard method is to use singular value decomposition (SVD), and keep the components that carry most variance. In our experience, 200 components are sufficient to capture >90% of the variance in the dataset and reduce the dataset a factor of 150[4,22]. A more specialized and efficient method for denoising and compressing widefield recordings is penalized matrix decomposition (PMD), which uses assumptions on the structure of noise for denoising and can achieve higher compression rates (~300 times reduction in size), with faster computation time[22].

The fourth step is to perform hemodynamic correction to remove non-neuronal related activity using measurements with violet light excitation. Before correcting, data are filtered with a 0.1 Hz temporal high-pass filter. An additional low-pass filter can be applied to remove fast transients in the violet channel since the non-neural signals are usually slower than neural signals. Frames acquired with violet illumination are then extrapolated to account for differences in acquisition time, rescaled to the amplitude of the frames acquired with blue illumination and subtracted from the frames with blue illumination. The result is dramatic for frames with high hemodynamic response where the blood vessels were clearly visible in frames acquired with violet and blue illumination but not in the corrected frames (Fig 3). Frames with no hemodynamic response are unaffected because the subtracted component is close to zero.

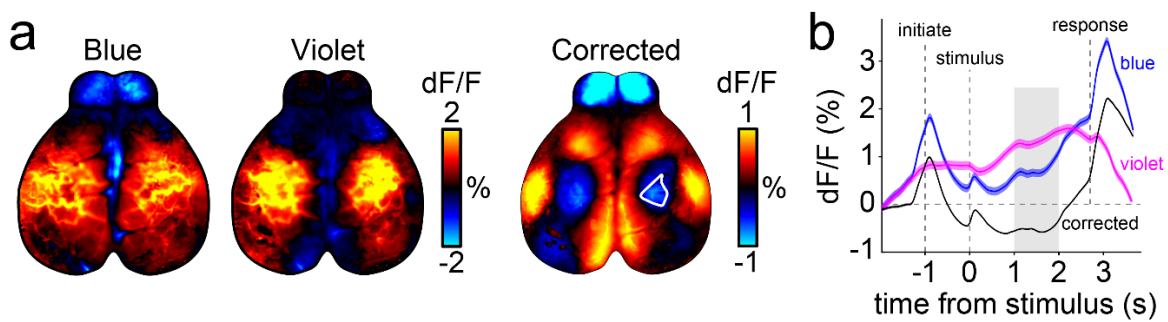

*Figure 3* **Correction of hemodynamic artifacts with alternating violet and blue illumination** a)Trial-averaged fluorescence with either blue- or violet excitation (left) during a behavioral task[4] (n = 402 trials). Hemodynamic corrected trial average (right) shows reduced activity in the hind-limb area (white outline). b) Hemodynamic correction recovers the temporal dynamics of neural signals. Traces show averaged fluorescence from the somatosensory hind-limb area (white outline in a)), which appears to be above baseline when averaging activity acquired with blue excitation. However, the hemodynamic corrected trace (black) shows that activity is below baseline. Gray shading shows the time from which maps in a) were averaged. Dashed lines show the time of different events in the task.

The last step is to extract features from the data that relate to measurable aspects of neural activity in distinct brain areas or cell populations, i.e. to interpret the data. Ideally, we would like signals to be referenced to well-defined brain regions. Averaging the activity of pixels could, in principle, be used to extract the activity of distinct brain areas. However, averaging can be problematic in presence of anatomical differences across animals and will not able to isolate co-activation of multiple brain regions. We propose LocaNMF, a method that decomposes widefield activity using existing brain atlases as a prior to obtain components localized to distinct cortical regions. This approach isolates signals from different areas and is capable of isolating overlapping signals in the same region (see Figure 4a in Saxena et al., 2020).

To streamline analysis and facilitate visualization of large datasets, we also developed tools for both Python (https://github.com/jcouto/wfield) and MATLAB (https://github.com/musall/WidefieldImager, also used in[4]) that are compatible with the aforementioned acquisition software, optimized to use limited computational resources and flexible to accommodate other imaging formats. Because longitudinal experiments generate large datasets that require a potentially expensive infrastructure for efficient data storage and processing, we also present an analysis pipeline that combines all the steps described above on the cloud-based analysis platform NeuroCAAS (http://neurocaas.org), which allows reproducible and streamlined analysis[40]. One of the major advantages of this platform is that it deploys well-established, state-of-the-art algorithms and runs on optimized hardware without local infrastructure and maintenance costs; users simply drag-and-drop datasets onto the system and then download the denoised, compressed, demixed output. Importantly, sophisticated analysis algorithms often take advantage of specific libraries or hardware that are often not trivial to implement locally and already implemented in the cloud. Using a cloud platform therefore streamlines data analysis, enables access to standardized high-performance algorithms, reduces processing time, eliminates hardware cost, and reduces the challenges of implementing processing pipelines on local premises. To simplify the interaction with the online platform, we created a graphical user interface, dedicated to launching analysis and retrieving results from the NeuroCAAS platform (https://github.com/jcouto/wfield).

**Limitations**

The depth of imaging through the intact cleared skull is limited by light scattering and therefore biased towards activity of superficial cortical layers. Placing the focal plane deeper into the cortex can blur the image but does allow more fluorescence from deeper cortical layers (Fig. 2d for example). In addition, novel molecular tools promise to overcome this limitation by expressing indicators exclusively in the somatic compartment[41]. This eliminates the contribution of dendritic or axonal processes to the emitted fluorescence. A key limitation is that widefield signals reflect the pooled activity over many neurons. Therefore, features encoded by the coordinated activity of different neurons in a local population and not reflected in the bulk activity of an area are elusive. Nonetheless, pooling activity over many neurons might be beneficial when measuring weak signals correlated across many neurons (e.g. inconspicuous sensory stimuli or arousal related activity).

The NeuroCAAS platform allows deploying a set of complex analyses on optimized hardware without requiring expertise in programing or hardware maintenance. However, since the service runs and stores data on Amazon's AWS cloud platform, it should be used with care when storing data with sensitive or confidential data. In addition, the underlying AWS services produce some costs when analyzing data; however the cost remains lower when compared to managing computational resources locally[40].

**Comparison to other methods**

Widefield imaging provides several advantages over conventional two-photon microscopy: large field of view, high temporal resolution (kilohertz range possible), low cost, and reduced complexity. However, these come at the expense of limited imaging depth and lack of single cell resolution. Another advantage over two-photon microscopy is the large working distance (~30mm) which makes widefield imaging easier to combine with electrophysiology[3,29]. Further, when compared to functional ultrasound and magnetic resonance imaging, widefield imaging provides higher temporal resolution and the possibility to target specific cell subpopulations.
The skull clearing technique has major advantages over cranial window preparations for longitudinal studies. Surgeries for implanting cranial windows require a high level of surgical proficiency, and optical clearance may degrade over time due to bone regrowth or infection. This limits the experimental throughput and increases the number of animals needed for a given study.

NeuroCAAS enables using recently developed algorithms for analyzing widefield data that complement and extend traditional methods of analysis on the cloud and aims to remove the burden of installing software packages that often have elaborate dependencies and/or acquiring and maintaining expensive computing hardware. In addition, it allows running analysis in parallel, which is a scalable alternative to running analysis sequentially on a single machine (where independent datasets cannot be processed simultaneously). An alternative to the cloud that is similarly scalable is to use a high-speed computing cluster. Due to the high cost of operating a high-speed computer cluster, some institutions operate and maintain clusters that make computing resources available to local personnel. However, these facilities can be costly to access and are available only to a subset of institutions with neuroscience laboratories. Finally, deploying analyses on a computer cluster often requires knowledge of how to operate the scheduling software that organizes computing resources, which can impose a barrier for some users and is greatly simplified in the cloud platform.

**Overview of the procedure**

There are four sections in this protocol as described above (Fig 1a). In Section 1, we describe how to assemble and calibrate the widefield macroscope (steps 1-24). In Section 2, we highlight our procedures for skull clearing (steps 25-40). In Section 3, we describe procedures for imaging neural activity in behaving mice expressing genetically encoded calcium indicators (steps 41-46). In the last section, we describe how to deploy analyses using the cloud-based NeuroCAAS platform (steps 47-49).

Our procedures can be used to probe the impact of behavior on neural activity across cortical areas, to establish the flow of activity in distant areas (e.g. in cortex-wide functional mapping experiments[14]), to perform longitudinal studies of cortical function, or to measure activity of distinct cortical cell-types by taking advantage of specific transgenic lines or injection strategies[42,43]). Widefield imaging can also be used in high throughput automated experiments[26,44] and to assess the impact of drugs on cortical activity or in disease models[45]. Finally, the protocol can be easily extended for high-speed imaging of voltage indicators[46], multi-color imaging of multiple fluorescent indicators[5,47] and for optogenetic manipulation[5,19].

**Materials**

**REAGENTS**

Transgenic mice expressing GCaMP6 in a neuronal population of interest (Charles River or Jackson Laboratories) **! CAUTION** Experiments involving animals must be conducted in accordance with relevant institutional and governmental guidelines and regulations.

**Skull clearing reagents**

Cyanoacrylate glue (Pacer Technologies, cat. no. ZAP-A-GAP CA+)

**Drugs**

Isoflurane (Covetrus, cat. no. 1169567761)

Ophthalmic ointment (Dechra, NDC 17033-211-38)

Ketamine (10%; WDT, cat. no. 793-319)

Meloxicam (5 mg/ml; Böhringer Ingelheim, cat. no. 141-219)

Lidocaine (1% (vol/vol); Bichsel, cat. no. 2057659)

**Other**

70% (vol/vol) Ethanol (EtOH)

Isotonic saline solution 0.9% (Ecoflac Plus container; B. Braun, cat. no.131321)

Black dental acrylic (Ortho-Jet, Lang Dental cat. no. 1520BLK and 1503AMB) – for bonding to the skull and the headbar and for light shielding.

**EQUIPMENT**

**Surgical Equipment**

Leica M60 stereomicroscope (Leica Microsystems, Model Leica M60)

Small rodent stereotaxic frame (Kopf, Model 900LS)

Fiber optic lamp with gooseneck (Schott, cat. no. SCHOTT-ACE-GOOSENECK)

Isoflurane vaporizer and anesthesia system (Kent Scientific, cat. no. VetFlo-1205S-M)

Feedback controlled thermal blanked (Kent Scientific, cat. no. PhysioSuite-PS-02)

**Surgical tools**

Scalpel (Fine Science Tools, cat. no. 10003-12)

Forceps (Fine Science Tools, cat. no. 11049-10)

Scissors (Fine Science Tools, cat. no. 14040-10)

Vannas scissors (World Precision Instrument, cat. no. 501777)

Micro probe, angled (Fine Science Tools, cat. no. 10032-13)

Delicate bone scraper (Fine Science Tools, cat. no. 10075-16)

Fine forceps (Fine Science Tools, cat. no. 11254-20, 11251-35)

High-speed dental drill (Foredom, cat. no. K.1070) – can be used to removed particles trapped in the cement

**Consumable supplies**

Syringes

Needles, 30G

Paper wipers

Cotton swabs

Sterile drape

Gauze

Gloves

Surgical mask

Hair net

Capillary glass pipette

Lens cleaning paper (Thorlabs MC-5)

**Software**

Matlab 2014b - image/data acquisition toolboxes

PCO frame grabber drivers (https://www.pco.de/fileadmin/user_upload/pco-driver/PCO_SISOINSTALL_5.7.0_0002.zip)

PCO Camware (https://www.pco.de/fileadmin/user_upload/pco-software/SW_CAMWAREWIN64_410_0001.zip)

PCO SDK (https://www.pco.de/fileadmin/user_upload/pco-software/SW_PCOSDKWIN_125.zip)

PCO matlab toolkit (https://www.pco.de/fileadmin/user_upload/pco-software/SW_PCOMATLAB_V1.1.2.zip)

Recording software - https://github.com/musall/WidefieldImager/tree/master/WidefieldImager

(Fully functioning Python alternative: https://bitbucket.org/jpcouto/labcams)

Preprocessing and graphical interface - https://github.com/jcouto/wfield

NeuroCAAS computing platform - http://neurocaas.org

**Macroscope** PCO edge 5.5 camera sCMOS, 2560X2160 with cameralink interface (PCO AG, cat. no. PCO.EDGE 5.5 RS AIR)

Nikon AF DC-NIKKOR 105mm f/2D lens (B&H, cat. no. NI1052DAF)

Rokinon 85mm f/1.4 AS IF UMC lens for Sony A (B&H, cat. no. RO8514S)

Nikon F to C Mount adapter (B&H, cat. no. VELACNF)

**! CAUTION** Parts numbers are for an imperial system, an alternative setup using metric parts is possible.

60 mm cage system removable filter holder (Thorlabs, cat. no. LCFH2 or omit if the filter is mounted directly on the DMF2)

2x Sensei 72-52mm step-down ring (B&H, cat. no. SESDR7252)

GCaMP excitation LED, 470 nm (Thorlabs, cat. no. M470L4)

Hemodynamic correction LED, 405 nm (Thorlabs, cat. no. M405L4)

2x T-cube LED driver (Thorlabs, cat. no. LEDD1B)

30 mm cage cube (Thorlabs, cat. no. C4W or DFM1 for easier assembly)

60 mm cage cube (Thorlabs, cat. no. LC6W or DMF2 for easier assembly),

3x Adapter with external SM2 threads and internal SM1 threads (Thorlabs, cat. no. SM2A6)

2" Optic mount for 60 mm cage cube with setscrew optic retention (Thorlabs, cat. no. LB5C1)

Fixed cage cube platform for C4W/C6W (Thorlabs, cat. no. B3C)

2x Adjustable collimation adapter (Thorlabs, cat. no. SM2F32-A)

30 mm cage compatible rectangular filter mount (Thorlabs, cat. no. FFM1)

2x Coupler with external threads, 1" long (Thorlabs, cat. no. SM1T10)

Lens tube spacer, 1" long (Thorlabs, cat. no. SM1S10)

2x SM1 lens tube, 1" thread depth with one retaining ring (Thorlabs, cat. no. SM1L10)

2x Light tight blank cover plate (Thorlabs, cat. no. LB1C)

Adapter plate for 1.5" post mounting clamp (Thorlabs, cat. no. C1520)

4x Coupler with external threads, 1/2" long (Thorlabs, cat. no. SM2T2)

Mounting platform for 60 mm cage cube, Imperial taps (Thorlabs, cat. no. LB3C)

Blue excitation bandpass filter, 470 nm with 40 nm bandwidth (Chroma, cat. no. ET470/40x) – to assert the wavelength of the blue LED to excite GCaMP

Violet excitation bandpass filter, 405 nm with 10 nm bandwidth (Edmund optics, cat. no. 65-133 or ET405/10x from Chroma) - to assert the wavelength of the hemodynamics correction LED
GFP bandpass filter, 525 nm with 45nm bandwidth (Edmund optics, cat. no. 86-963 or ET525/36m from Chroma (quote for 50mm)) – for restricting the light that reaches the camera sensor
Dichroic mirror, R: 325-425, T: 444-850, 435 nm cut-off, 25.2mm x 35.6mm, 1.05mm thick (Edmund optics, cat. no. 87-063) – intended to combine the excitation light, lets pass blue and reflects violet
495 nm long-pass dichroic mirror, R: 450-490, T: 500-575 and 600-700nm, 50mm diameter, 1mm thick (Chroma, cat. no. T495lpxr, ring mounted – use T495LPXR-UF2 if a rectangular filter is needed as when using the DMF2) – to reflect blue and violet light to the sample and let pass green fluorescence to the camera sensor

Post mounting clamp (Thorlabs, cat. no. C1511)

Mounting base (Thorlabs, cat. no. BA2)

3x Mounting posts (Thorlabs, cat. no. P8)

4x Optical posts (Thorlabs, cat. no. TR8)

Optical breadboard 36" x 72" x 2.28" with 1/4"-20 mounting holes (Thorlabs, cat. no. B3672FX)

Alternative for a smaller setup: Optical Breadboard, 36" x 48" x 2.28" with 1/4"-20 mounting holes (Thorlabs, cat. no. B3648FX)

High rigid frame 900 mm x 1500 mm (3' x 5') (Thorlabs, cat. no. PFR90150-8)

Alternative for 36" x 48" breadboard: High rigid frame 750 mm x 900 mm (2.5' x 3')  (Thorlabs, cat. no. PFR7590-8)

Lab Jack (Thorlabs cat. no. L490) to adjust the height of the behavioral setup or focussing module (Thorlabs cat. no. ZFM1020)

Power meter (Thorlabs, cat. no. PM100D)

Photodiode power sensor, 400 - 1100 nm (Thorlabs, cat. no. S120C)

Teensy 3.2 (PJRC, cat. no. TEENSY32)

SMA female RF coaxial adapter PCB mount plug (Amazon, ASIN B06Y5WZ1TK)

USB to micro USB, 6ft (Digikey, cat. No. AE10342-ND)

Prototyping perforated breadboard 2" x 3.2" (Adafruit, cat. no. 1609)

BNC connector jack, female socket 75 Ohm panel mount (Digikey, cat. no. A97562-ND)

Adapter coaxial connector, female socket to BNC jack (Digikey, cat. no. ARFX1069-ND)

Coaxial BNC to BNC male to male RG-58 29.53" (750.00mm) (Digikey, cat. no. J10339-ND)

Coaxial SMA to SMA male to male RG-316 5.906" (150.00mm) (Digikey, cat. no. J10284-ND)

BNC, Male plug to wire lead 72.0" (1828.80mm) (Digikey, cat. no. 501-2556-ND)

3D-printed light shielding cone ('Polished Metallic Plastic', Shapeways.com).

CAD files available in https://github.com/musall/WidefieldImager.

Data acquisition computer; at least 4-core CPU with 32GB RAM; with a full-length PCI slot for the frame grabber and a separate solid state drive for data acquisition.

**REAGENT SETUP**

**! CAUTION** Prepare reagents using aseptic techniques. Some of the reagents are controlled substances.

**Ketamine/ Dexmedetomidine anesthetic**

Mix 8.4 ml of 0.9% NaCl with 0.6 ml of Ketamine (from a 100 mg/ml stock) and 1 ml of Dexmedetomidine (from a 0.5 mg/ml stock) in a sterile vial. Inject 0.3 ml for a 30 g mouse to deliver a dose of 60 Ketamine / 0.5 Dexmedetomidine (mg/kg).

**Meloxicam**

Mix 9 ml of 0.9% NaCl and 1 ml of Metacam (5 mg/ml stock) into a sterile vial.

**EQUIPMENT SETUP**

The implants for head fixation (headpost) were laser cut from a 0.05" Titanium sheet. An aluminum custom-machined holder secures the headpost during imaging (CAD files in https://github.com/musall/WidefieldImager).

**Procedure**

**Macroscope assembly** TIMING 2-3 h

Clear the optical table, and gather the tools and parts for the build.

**Steps 1-4: Assemble the 60mm cage cube** – Fig. 2a TIMING 20 min

**1|** Attach the optic mount (Thorlabs, LB5C1) to the mounting platform (Thorlabs, LB3C). Place the 495 nm long-pass dichroic mirror (Chroma, T495lpxr) in the optic mount.

**CRITICAL STEP** Orient dichroic correctly. The correct orientation is in the datasheet that comes from the manufacturer. When handling optics use gloves and lens paper, and be careful not to scratch the optics.

**2|** Tighten two couplers (Thorlabs, SM2T2) to the cage cube (Thorlabs, LC6W).

**3|** Drill four holes (slightly larger than the 8-32 screws) on a cover plate (Thorlabs, LB1C) – these will attach to the post clamp (Thorlabs, C1511) see Fig. 2a. Use a workshop (e.g. from your institute) if possible. Cover the side of the cube (that will be used for supporting the macroscope, see Fig. 2d) with the machined cover plate (Thorlabs, LB1C). Attach an additional cover plate (Thorlabs, LB1C), orthogonal to the machined cover plate, on the cage cube.

**4|** Attach the SM1 to SM2 adapter (Thorlabs, SM2A6) to the side directly facing the dichroic.

**Steps 5-10: Assemble the 30 mm cage cube** – Fig. 2 b TIMING 30 min

**5|** Attach the rectangular optic mount (Thorlabs, FFM1) to the mounting platform (Thorlabs, LB3C). Secure the excitation dichroic mirror (Edmund, 87-063) on the optic mount.

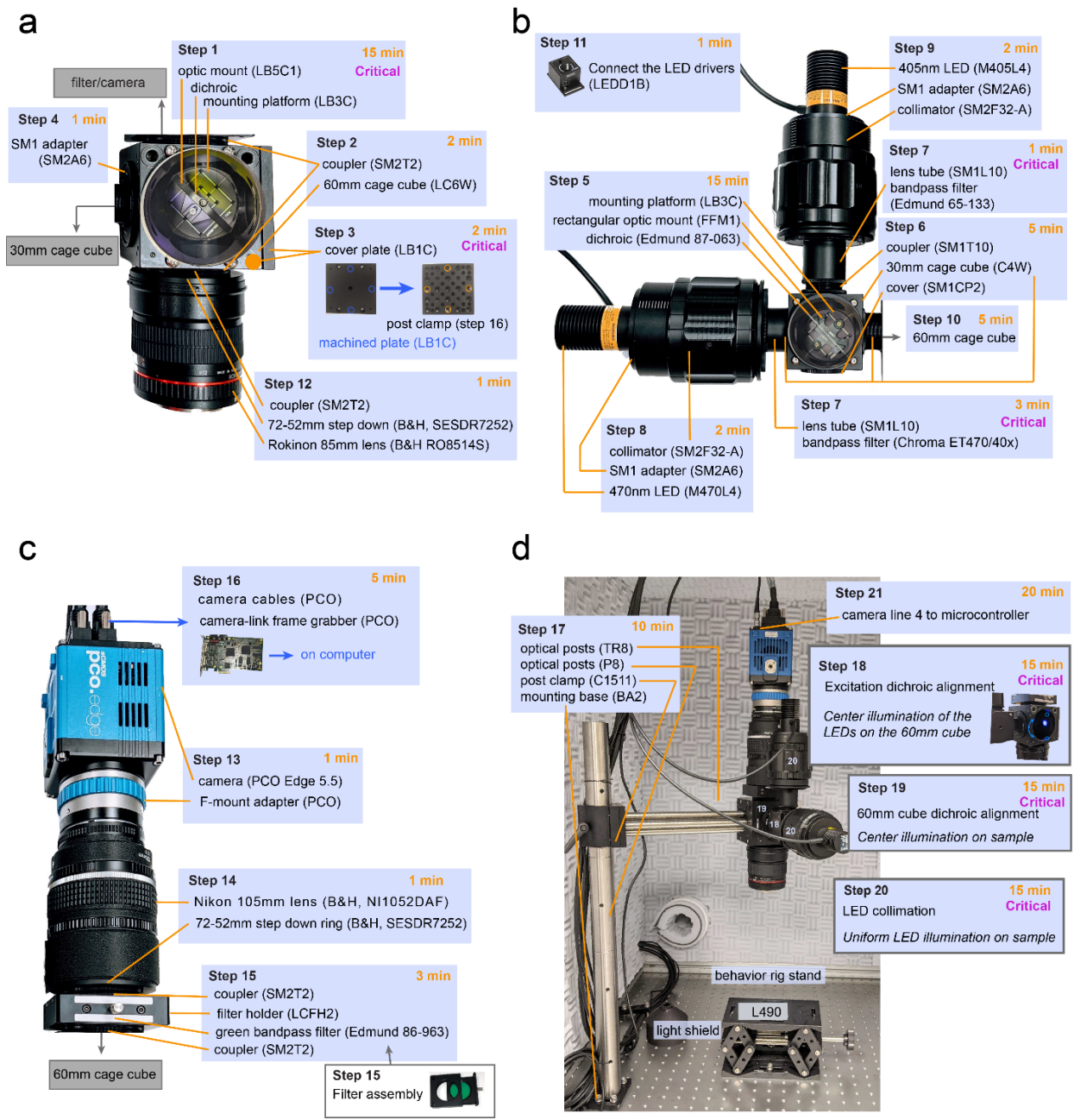

*Figure 4 **Detailed build instructions for the widefield macroscope**. a) Steps for assembling the main cage cube and objective module. b) Steps for assembling the excitation cage cube and light collimation module. c) Details for assembling the light collection module. d) Example setup with a rig stand and highlighting the steps for calibration.*

**6|** Attach two SM1 couplers (Thorlabs, SM1T10) to the cage cube (Thorlabs, C4W). Attach an SM1 end cap (Thorlabs, SM1CP2) to the bottom of the cube. Attach the assembly from step 5 to the cage cube. Seal the cage cube with a blanking cover plate (Thorlabs, B1C) on the side opposing the dichroic mirror.

**7|** Secure the violet bandpass filter (Edmund 65-133) to a lens tube using the retainer rings (Thorlabs, SM1L10). Attach the lens tube to the top of the cage cube. Repeat for the blue bandpass filter (Chroma ET 470/40x) and connect the assembly to the side of the cage cube. **CRITICAL STEP** Follow the filter manufacturer's instructions to orient the filter correctly. Install the LEDs so that violet light is reflected and blue light passes through the dichroic mirror.

**8|** Connect the 470 nm excitation LED (Thorlabs, M470L4) to one of the collimators (Thorlabs, SM2F32-A). Attach the whole assembly to the blue filter lens tube using an SM2 adapter (Thorlabs, SM2A6).

**9|** Similar to step 8, connect the 405 nm LED (Thorlabs, M405L4).

**10|** Connect the 30 mm cube assembly to the larger 60 mm cube, by connecting a lens tube spacer (SM1S10) to the SM2-adaptor ring (Thorlabs, SM2A6).

**11|** Connect the LEDs to the T-cube LED drivers (Thorlabs, LEDD1B) and set the drivers current limit to the maximum (arrow on the side of the driver) using a screwdriver. **CRITICAL STEP** The maximum driver current is 1.2A but the LEDs are rated for 1A. In our experience, using 1.2A works and provides more power but might reduce LED lifetime.

**Steps 12-16: Connect the lenses, camera, and green filter** TIMING 10 min

**12|** Attach a 72-52 mm step-down ring (B&H, SESDR7252) to the Rokinon 85 mm lens (B&H, RO8514S) with the lens is in an inverted position (the side typically facing the camera is facing the object in this configuration). Use a coupler (SM2T2) to secure the step-down ring and lens to the bottom of the cage cube (Fig. 2a). **CRITICAL STEP** Using a lens with different threading may require a different step-down ring.

**13|** Attach the Nikon F to C mount adapter (B&H, VELACNF) to the PCO Edge camera.

**14|** Secure a 72-52 mm step-down ring (B&H, SESDR7252) to the Nikon 105 mm lens (B&H, NI1052DAF). Attach the Nikon 105 mm lens (B&H, NI1052DAF) to the f-mount adapter and lock it in place. **CRITICAL STEP** Like for step 12, a different lens may require a different step-down ring.

**15|** Prepare the filter assembly. Secure the green bandpass filter (Edmund 86-963) to the filter holder (Thorlabs, LCFH2). Attach two couplers (Thorlabs, SM2T2) to the filter holder cage. Connect the holder to the step-down ring on the camera assembly and to the top of the 60mm cage cube (Fig. 2a).

**16|** Install the frame grabber in the acquisition computer. Turn ON the computer and install the camera drivers and software following the camera manufacturer instructions (PCO frame grabber drivers, PCO Camware and PCO SDK). Connect the cables from the camera to the computer. Connect the power cable of the camera.

**Step 17: Secure the widefield assembly to the optical table TIMING 10 min**

**17|** Combine 3 large optical posts (Thorlabs, P8), attach them to the mounting base (Thorlabs, BA2) and secure the assembly to the optical table. Attach 4 optical posts (Thorlabs, TR8) to the machined cover plate (from step 3). Secure the optical posts to the post clamp (Thorlabs, C1511). Slide the full assembly on the 1.5"optical posts (Thorlabs, P8).

**Steps 18-20: Calibrate the excitation light path TIMING 45 min**

**18|** Loosen the setscrew on the collimators (from step 9 and 8). Remove the dichroic on the 60 mm cage cube (from step 1). Loosen the cover on the 30 mm cage cube (from step 5) enough to rotate the mirror. Turn on the blue and violet LEDs, use a moderate power level to be able to see both. Adjust the collimators to obtain a small square image of each LED on the back of the 60 mm cube. Align the violet light on the blue LED by rotating the dichroic from step 5 (Fig. S2a). Secure the dichroic in place when the violet and blue LEDs directly on top of each other. Put the dichroic back in the 60 mm cage cube. **CRITICAL STEP** Note that this step is not required if using a DFM1 cage cube that also allows quickly replacing the dichroic. However, alignment between the two excitation LEDs is a critical step that if done improperly, will impair the hemodynamic correction.

**19|** Turn the camera ON and open the Camware software in the computer. Click the preview button. Activate the grid in the software using the mouse dropdown menu. Open the aperture of both lenses completely (rotate clockwise). Place a white paper under the Rokinon lens (i.e. the objective) at focus height. Loosen the dichroic in the 60mm cage cube. Turn on the blue LED until it almost saturates the center of the preview image (Fig. S2b). Rotate the dichroic to center the light in the camera image. Secure the dichroic in place by tightening the screws. **CRITICAL STEP** This step is essential to illuminate the sample evenly. Note that this step is not required if using a DFM2 cage cube because the mirror is fixed at 45 degrees in the kinematic filter mount.

**20|** On Camware activate the line profiler on the mouse dropdown menu. Select a line that covers the camera sensor (Fig. S2c). Adjust the collimators for each LED until the light is uniformly covering the entire preview image and the profile histogram is flat on both channels (Fig. S2c). Use a power meter (Thorlabs, PM100D; probe S120C) to measure the power at the

sample (470 nm ~160 mW with 1.2A; 405 nm ~44 mW with 1.2 A). **CRITICAL STEP** This step is important to achieve even illumination of the sample.

**Steps 21 and 24: Synchronize the illumination and the camera TIMING 25 min**

**21|** Connect line 4 of the camera (all lines exposure signal) to the microcontroller (PJRC; Teensy 3.2) circuit (pin 3). Connect the LED drivers to pin 5 (470 nm LED) and 6 (405 nm LED) (Fig. 2b and c). Connect the microcontroller to the computer and load the Teensy firmware (LEDswitcher from the WidefieldImager repository) using the Arduino IDE.

**22|** Connect the DAQ (National Instruments; USB-6001) to the acquisition computer. Connect the Teensy microcontroller and the digital and analog lines as shown in Fig. 2c. **CRITICAL STEP** The trigger signals control the start and end of each trial. Analog input ports are used to synchronize with behavior offline.

**23|** Set the LED drivers to triggered mode (switch in the middle position). Configure camera output sync in the Camware software: Go to *Features | Hardware IO control | Status Expos | Signal: Status Expos; Timing: Show common time of all lines; Status: ON; Signal: High.*

**24|** Clone or download the github repository https://github.com/musall/WidefieldImager and add it to the MATLAB path. Install the pco.matlab adaptor package (PCO, check materials). Type 'imaqhwinfo' in MATLAB. The camera should appear as 'pcocameraadaptor'. Type 'WidefieldImager' in the command window and click 'start preview' to see a video stream from the camera. Click 'Light OFF' to switch on the blue LED.

**Skull clearing TIMING 1 - 2 h**

**CRITICAL STEP:** Apply aseptic techniques for this step.

**25|** Place the animal in the induction chamber, prefilled with Isoflurane (3% in 100% oxygen 0.5-1 L/min) for 20-60 sec. This step reduces stress and facilitates handling.

**! CAUTION** All experiments involving animals must be performed in accordance with relevant institutional and governmental guidelines and regulations.

**26|** Induce anesthesia with an intraperitoneal injection of Ketamine-Medetomidine (60/0.5 mg/kg). Let the anesthesia take full effect (usually 5-10 min). Confirm the anesthesia depth with the absence of toe-pinch reflexes. The procedures can only start when the mouse is deeply anesthetized.

**27|** Place the animal in the stereotaxic frame, making sure that the earbars and the incisors are firmly holding the skull. Make sure that the feedback-controlled heating blanket is operating and maintaining body temperature.

**28|** Inject Meloxicam (1-2 mg/kg) subcutaneously in the back of the animal for systemic analgesia. Apply Lidocaine on the scalp for topical analgesia.

**29|** Cover the eyes with ophthalmic ointment to prevent drying.

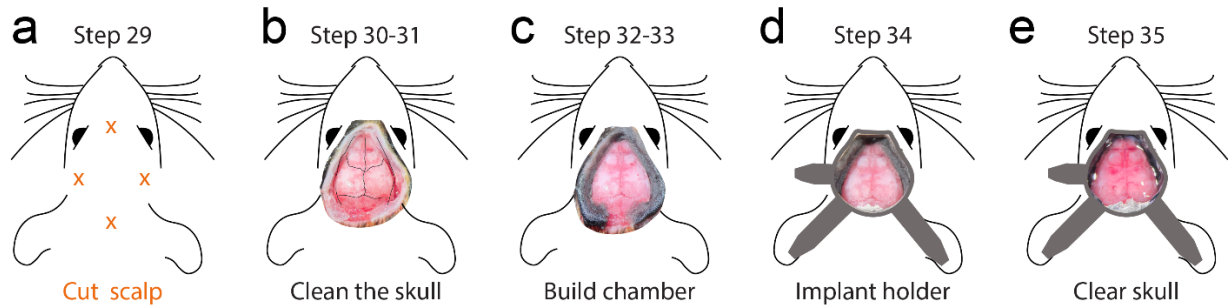

*Figure 5* **Overview of the skull clearing procedure.** *a) Location of scalp cuts in step 30. b) Example of clean skull preparation with temporal muscles removed in step 32. c) Example of the chamber in step 34. d) Placement of the headpost implant for head restraint (step 35). e) Example of a mouse with the skull cleared in the end of the surgery (step 36).*

**30|** Trim the hair in a region starting 2 mm in front of the eyes until the middle of the ears (anterior-posterior) and below the level of the eyes (lateral). Use a pair of scissors or alternatively an electric trimmer. Apply a hair removal product (e.g. Nair) if needed to remove all hair from the surgical area.

**31|** Clean the area using cotton swabs alternating between 70% ethanol and betadine.

**32|** Make 4 small incisions using a fine pair of scissors in the locations highlighted in Fig. 4a. Remove the scalp using the scissors to cut between the incisions and a pair of forceps.

**33|** Use the number 11 scalpel blade or a pair of fine forceps to push the periosteum to the edges of the skull. Scrape any remaining periosteum until the skull is clean. **CRITICAL STEP** This step is essential for proper optical clearance. It is critical that there are no scratches or damage to the skull.

**34|** Retract the temporal muscles slightly using a pair of closed forceps. Use the same approach to carefully retract the muscles posterior to the lambda suture.

**35|** Apply VetBond to the exposed skin 1mm around the cuts (Fig. 4b).

**36|** Apply dental cement to cuts and in the temporal areas. The cement should form a flat surface to which the headpost can be attached (Fig. 4c).

**37|** Place the headpost and hold it in place with vetbond and dental cement (Fig. 4d).

**38|** After the cement is cured, apply cyanoacrylate glue to the skull. Start at the sutures and cover the entire skull. Only apply a small amount of cyanoacrylate and form a thin layer over the skull. Then wait for the layer to cure and repeat the procedure 2-3 times. Use a toothpick to distribute the cyanoacrylate. **CRITICAL STEP** The cyanoacrylate cures spontaneously within minutes. We found that repeated application of 4-5 layers of cyanoacrylate produces the best results. It takes ~5 minutes for the first layer cure. Subsequent layers tend to cure more quickly. Be careful because it can cure faster it makes contact with the dental cement. Too fast curing from the sides can cause heat which is harmful to the clearing. To avoid this, build the first layer from the center and avoid touching the cement walls. Fill gaps on the sides when adding subsequent layers.

**39|** Wait until the glue is dry. Skull clearing should set in gradually and brain vessels should be clearly visible after 5-10 minutes. Inject Antisedan to reverse anesthesia and monitor the animal as it recovers.

**40|** Monitor the animal recovery for at least 48h and inject analgesia as needed (Meloxicam every 24h). **CRITICAL STEP** While optical clarity settles less than one hour after the procedure, allow enough time for recovery and habituation before imaging to minimize animal stress.

**Imaging TIMING 1 - 2 h**

**41|** Open the Camware software. Go to the *Camera settings* tab and select *I/O Signals*. On *Line 4* set *Signal* to *Status Expos.* Click the clock icon and select "*Show common time of 'All Lines'*". **CRITICAL STEP** This step is critical to avoid artifacts because of the moving shutter. The camera setting ensures that the exposure trigger is only active when all lines of the camera is exposing. Cameras with global shutter allow all lines to be read simultaneously but usually do not allow high sampling rates.

**42|** Type the command "WidefieldImager" in the MATLAB command prompt to start the acquisition software (Fig. 2d). Click "New animal" to create a new animal ID and enter a name for the current experiment in the pop-up dialog. Make sure other settings, such as the trial duration and file path are correct. A detailed, step-by-step instruction manual that includes descriptions of all features and further details for how to perform an imaging session with the software is in https://github.com/musall/WidefieldImager.

**43|** Attach the animal to the behavior apparatus. Clean the implant with a cotton swab dipped in ethanol. Click "start preview", center the animal under the objective and switch on the blue light. Position the light shielding cone over the implant. **CRITICAL STEP** Make sure the shielding blocks all the light but does not push on the animals' eyes nor occludes the field of view. The light shield is essential because the camera operates in rolling shutter mode. It may be omitted, for experiments in absolute darkness. However there may be activation in visual cortex upon light onset because mice will be able to see the excitation light.

**44|** Open the aperture of the imaging lens completely (Fig. 1c) and adjust the height of the setup until the most lateral brain vessels are clearly in focus. Due to the brain curvature, the center of the implant will appear blurry (Fig. 2d). Adjust the light power of the blue and violet LED. **! CAUTION** More excitation light results in stronger fluorescence signals but too much light will saturate the camera or cause photobleaching of the indicator (Fig. S1). We found that light power between 10-25 mW is an appropriate range but the optimal light intensity strongly varies, depending on the localization, expression level and brightness of the indicator. To avoid saturation, we recommend setting the light levels so that the brightest pixels remain below 70% of the cameras full dynamic range. Pixels above this threshold are shown in red in the camera preview window (Fig. 2d). To avoid photobleaching, the light intensity should remain below 50 mW (Fig. S1). For dim samples, reduce the framerate to increase the exposure time and therefore the measured signal.

**45|** Focus on the blood vessels over the cortical surface (Fig. 1d) and click "Take Snapshot" to save this image to disc for later reference. The depth of field can be extended

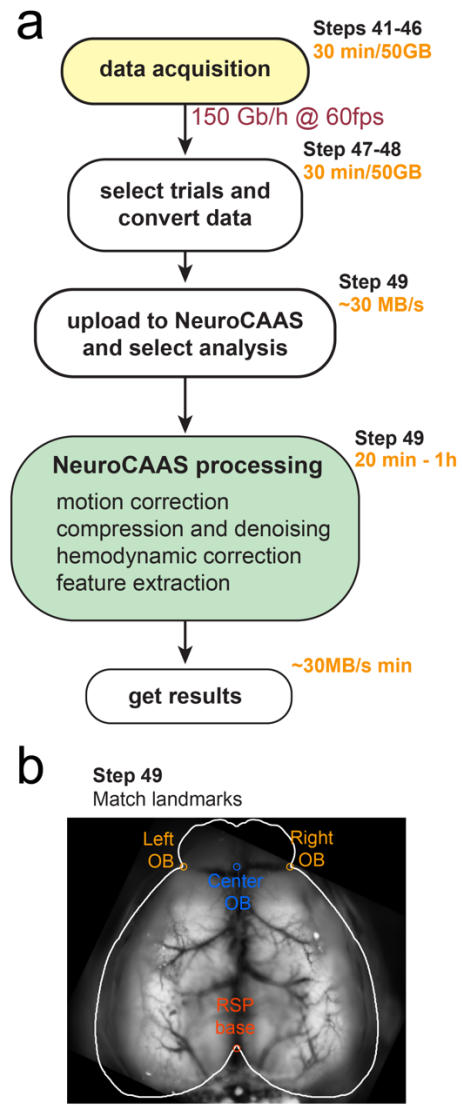

*Figure 6 **Diagram of the data processing pipeline.** a) Following data acquisition, data from all trials are concatenated (step 47) and uploaded to the cloud platform - NeuroCAAS (step 49). Preprocessing steps are performed in the cloud (step 49) and data are retrieved for further analysis (step 50). b) Landmarks for roughly match to the Allen Common Coordinate Framework (Step 48). This step is required for using locaNMF.*

momentarily to obtain a sharper image of the blood vessels by closing the aperture. To ensure that a blood vessel image is acquired in every session, the recording cannot be started before an image is taken. Make sure that the aperture is fully open. Focus on the most lateral brain vessels again, set excitation light to "Mixed light" and click the "Wait for trigger" button. The software is now ready to acquire imaging data and waiting for a "trial start" trigger. The trigger needs to be sent to the NI-DAQ to start acquisition (Fig. 2b). If the DAQ is not connected, you can mimic triggers in software by clicking the "Trial Trigger" button.

**46|** Start the stimulation or behavioral paradigm and send a "trial start" TTL trigger to the DAQ. The software starts acquiring frames and waits for a "stimulus trigger" TTL. When a "stimulus trigger is received, the software will record the baseline frames and acquire frames until the end of the post-stimulus period. If no stimulus trigger is received within the "wait" period no data are recorded, the trial is stopped and a new "trial start" trigger needs to be provided. An optional "trial end" trigger can be provided to stop a trial at any time. Monitor the MATLAB command window to ensure that the correct number of frames were saved in each trial.

To stop the recording, unlock the recording panel by clicking click the "Locked" button and end the current session by clicking the "Wait for Trigger" button.

**Data pre-processing TIMING 20 min - 4 h**

**47|** Follow the instructions to install the processing tools https://github.com/jcouto/wfield#installation. Open a terminal and type **wfield imager <path to data folder> -o <converted folder>** to concatenate data from all trials.

**48|** In the terminal run: **wfield open_raw <converted folder>.** To roughly match each imaging session to the Allen Common Coordinate Framework, drag the points from the left, center and right olfactory bulb as well as the base of the retrosplenial cortex to the respective landmarks. Press the **save** button to store the results. This step generates a transformation matrix to align the imaging data to the Allen Common Coordinate Framework and is required for some algorithms/analysis (e.g. locaNMF).

**49|** Run the analysis through NeuroCAAS either using the graphical user interface or uploading data to the website (http://neurocaas.org/). The website provides user interfaces for Penalized Matrix Decomposition and LocaNMF. To run the analysis from the web interface go to http://neurocaas.org and create user credentials (use group code EF1E04 for testing). Log in and navigate to the "Available analysis" page. Click on "View analysis" followed by "Start analysis". Select a dataset by clicking the checkbox next to the files or drag and drop your own

dataset to upload. Tune analysis parameters in the configuration file by downloading, changing the parameters value with a text editor and re-uploading. Click on a config file to select the parameters. Start the analysis by pressing the "Submit" button. A dialog will appear when the analysis completes (in under 15 min for the example datasets) after which the results can be downloaded.

Alternatively, you can run analysis from the graphical user interface. This can be used for motion correction, Penalised Matrix Decomposition, hemodynamics compensation and LocaNMF. Open a terminal and type the command **wfield ncaas**. The first time, it will ask for the credentials. Close the dialog window after inserting the neurocaas.org credentials. Drag the data folder from the local computer browser (on the left) to the cloud browser window (on the top right). On the dialog window, select the analysis to run, choose the analysis parameters and select the files to upload. Press "Submit" to close the dialog and add the analysis to the transfer queue. To upload data and start the analysis press the "Submit to NeuroCAAS" button. Monitor the analysis progress by clicking on the log files that appear on the cloud browser. Instructions for operating the graphical interface are in https://github.com/jcouto/wfield/blob/master/usecases.md#using-the-wfield-ncaas-interface. The results of preprocessing, i.e. the motion corrected, denoised, compressed, and demixed data, are copied back to the local computer when the analysis completes. **CRITICAL STEP** An account on NeuroCAAS is required to run the analysis. For the compression step, the values for block height and width depend on the frame size and must allow dividing the frame in equal chunks (e.g. for a frame height of 540, we use block height 90; and for frame width 640, block width 80), this needs to be set in the parameters.

**Timing:**

Steps 1-24, Macroscope assembly and alignment: 2-3h

Steps 25-40, Surgery: 2h per mouse

Steps 41-46, Imaging: 1-2h per session

Steps 47-49, Data preprocessing: 15 min to 3 hours depending on the dataset and the analysis

**Troubleshooting table:**

| Step | Problem | Possible reason | Solution |
|---|---|---|---|
| 17 | Lenses wobble | Couplers are not tight. | Readjust couplers in steps 13-15. |
| 19 | No light out of the objective | Miss-aligned dichroics, incorrect cube orientation | Check the orientation of the filters in step 7 and the dichroics in step 1 and 5. |
| 18 | Illumination is not even across channels | Improper alignment of the dichroic in the excitation cube. | Repeat the alignment step 18. |
| 21 | The object is not fully iluminated | The aperture of the objective (Rokinon lens) is partially closed or the LEDs not collimated. | Rotate the aperture of the objective clockwise. Repeat step 20. |
| 24 | Imager can't find "Dev1" | DAQ numbering not correct. | Use the command daqlist('ni') in MATLAB to discover the name of the USB DAQ device. Edit the daqName in line 63 of WidefieldImager.m. |
| 40 | Slow recovery after surgery, the animal seems irritated by the implant. | Excessive dental cement around the implant. | Use less dental cement on the sides of the implant, avoid sharp edges or open pockets that may cause skin irritation or infection. |
| 40 | Mouse is stressed in the setup or has poor task performance | Headpost too close to the eye or light shield is miss placed. | Make sure that the implant is at least 3mm from the eye by creating a structure from dental cement before securing the headpost (step 31-34). |
| 39 | Skull clearing is uneven or seems opaque | Applied glue before letting the cement cure. | Let the cement cure before applying the cyanoacrylate. If the glue contacts the liquid phase of the cement, it can create uneven patterns. |
| 39 | Cyanoacrylate is uneven or there are air bubbles | Too much glue or application too fast. | Apply thin layers of cyanoacrylate at a time and allow time to cure. |
| 40 | Dirt trapped in the cyanoacrylate after clearing or dents or cracks appear after weeks. | Short curation time. | Allow more time for the cyanoacrylate to cure before finishing the procedure. Use a dental drill to remove the debris and re-apply the cyanoacrylate. |
| 44 | No signal after repeated imaging | Photo-bleaching | Increase the interval between consecutive imaging sessions, reduce LED power, and/or decrease the acquisition rate. |
| 44 | The imager does not show a preview of the camera | The software could not connect to the camera | Make sure the camera is accessible through Camware. Close Camware and attempt restarting the Imager. Restart the computer and check the driver installation if the problem persists. |
| 44 | Error in the number of frames collected | The acquisition settings are incorrect. | Check the "framerate", "baseline" and "poststim" duration settings in the Imager. |

**Anticipated results:**

This procedure allows functional imaging of the entire dorsal cortex through the intact skull of awake mice. There are many possible applications for functional imaging: mapping experiments can identify the location of many sensory areas, based on their neural response to repeated visual or somatosensory stimulation (Fig. 6a and b). By careful quantification of animal behavior, a similar approach can be used to identify areas that are related to specific animal movements (Fig. 6c). Responses to sensory stimulation or movements can also be extracted when mice perform a behavioral task. For example, a linear encoding model can be used to isolate cortical activity patterns that are most related to different task or movement events[4,8,38] (Fig. 6d). Because the procedure is minimally invasive and leaves the skull intact, mice can be imaged for long timespans (we confirmed stable implants up to 1 year). This makes widefield imaging ideal to observe changes in cortical activity during task learning[48] (Fig. 6e) or for following long-term changes due to perturbations such as a stroke[45] or tumor growth[49]. Highly specific transgenic mouse lines enable imaging genetically defined neuronal sub-populations[43] (Fig. 6f) and provide exciting new possibilities when combined with whole-cortex widefield imaging.

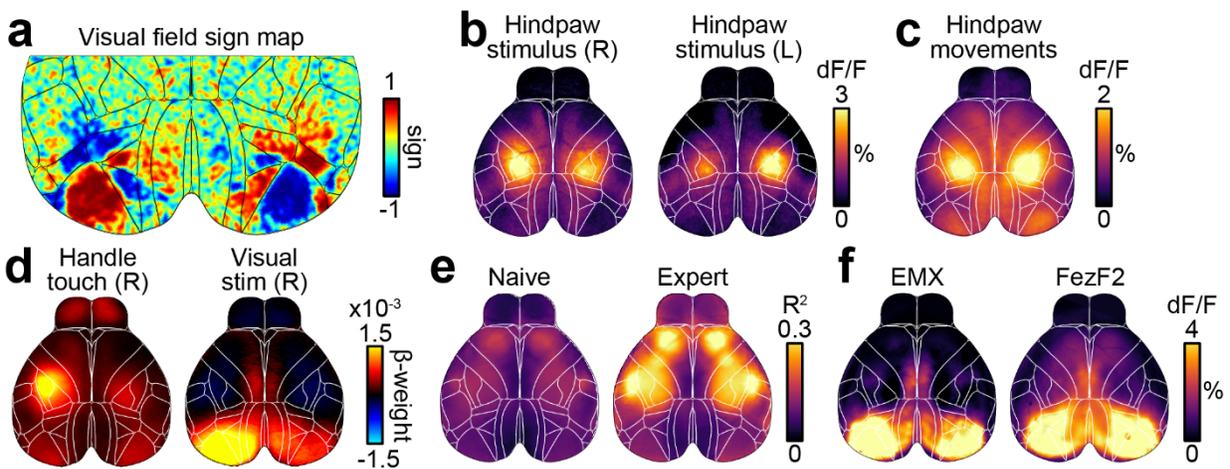

*Figure 7 **Widefield imaging applications.** a) Retinotopic mapping reveals the location of primary and secondary visual areas in both hemispheres. b) Somatosensory hindpaw stimulation shows robust and localized activation of somatosensory cortex. c) Analyzing data from a pressure sensor can be used to infer hindpaw movements. Cortical activation strongly resembles sensory stimulation data in b). d) Model weights from linear regression analysis can show cortical response patterns to different variables, such as touching a handle or visual stimulation. e) Example data from two recordings in the same mouse, 6 weeks apart. Imaging quality is stable and task-related cortical activity strongly increases with learning. f) Imaging specific cell populations using transgenic mouse lines. Average dF/F during visual stimulus presentation for a mouse with expression in all pyramidal neurons (controlled by the emx1 gene) or restricted to a subpopulation of layer 5 pyramidal neurons (FezF2). Despite expression being restricted to deep layers, the response to the visual stimuli in the FezF2 mouse line is comparable to the activity recorded in an Emx1 line.*

Taken together, this protocol puts within reach of many investigators a powerful platform for collecting, processing and analyzing cortex-wide responses in awake animals. Compared to other imaging approaches, the reduced cost, streamlined assembly and integrated analysis tools that define our platform will equip diverse investigators to make new discoveries about brain function.

**Author contribution statement**

S.M. and A.K.C. conceptualized early versions of the procedures. S.M. and S.G. implemented early setup versions and acquisition workflow. S.M., X.R.S. and J.C. refined surgical procedures. All authors refined the macroscope building procedures and compiled the required part lists. S.M. and J.C. wrote acquisition software. S.M., X.R.S. and S.G. prepared animals and acquired data in the expected results. J.C., S.M., I.K. and S.S. wrote software for analysis. T.A., I.K., S.S., J.P.C. and L.P conceptualized the general analysis workflow. T.A., I.K. and S.S. deployed software on NeuroCAAS with input from J.C. and S.M.. A.K., J.C. and S.M. prepared figures. J.C., S.M., S.G., A.K. and A.K.C. wrote the manuscript with input from all authors.

**Acknowledgments**

We thank Matthew Kaufman and Kachi Odoemene for help with developing early versions of the protocol. Priyanka Gupta, Florin Albeanu and Joe Wekselblatt for technical advice. Nick Steinmetz, Marius Pachitariu and Kenneth Harris for help with widefield analysis.
Financial support was received from the Swiss National Science foundation (S.M., grant no. P2ZHP3_161770), the NIH (grant no. EY R01EY022979 and BRAIN initiative 5R01EB026949) and the Army Research Office under contract no. W911NF-16-1-0368 as part of the collaboration between the US DOD, the UK MOD and the UK Engineering and Physical Research Council under the Multidisciplinary University Research Initiative (A.K.C.). X.R.S. was supported by the NINDS BRAIN Initiative of the National Institutes of Health under award number F32MH120888. T.A. was supported by NIH training grant 2T32NS064929-11. S.S. was supported by the Swiss National Science Foundation P400P2 186759 and NIH 5U19NS104649. J.P.C. was supported by Simons 542963 and the McKnight Foundation. L.P. was funded by IARPA MICRONS D16PC00003, NIH 5U01NS103489, 5U19NS104649, 5U19NS107613, 1UF1NS107696, 1UF1NS108213, 1RF1MH120680, DARPA NESD N66001-17-C-4002 and Simons Foundation 543023. L.P. and J.P.C. were supported by NSF Neuronex Award DBI-1707398.

**Competing interests**

The authors declare no competing interest.

## Supplementary figures

### Figure S1 – Photobleaching

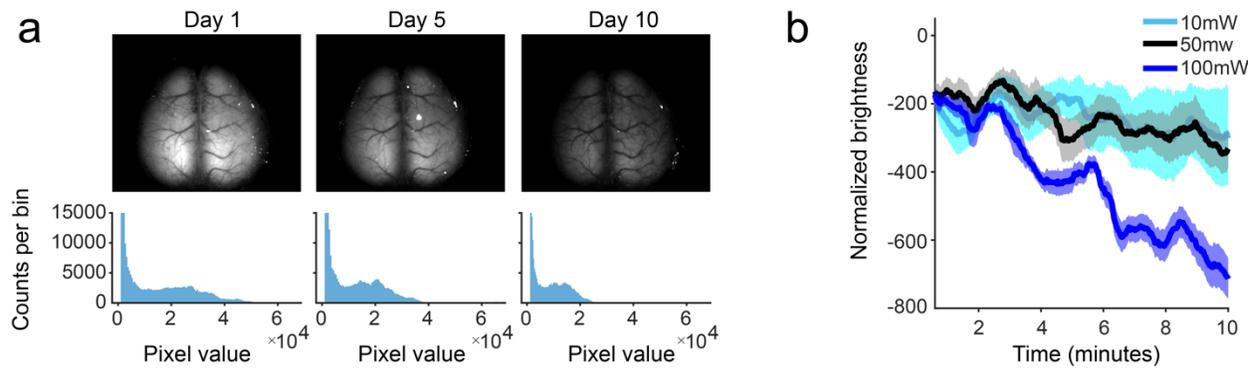

*Figure S1 – **Photobleaching due to repeated imaging or high power** a) Photochemical degradation of the fluorophore can occur over days because of repeated imaging. (a) Top: images from a mouse expressing GCaMP6 in a subpopulation of cortical excitatory projection neurons. Visible decrease in overall fluorescence is evident after daily imaging at day 5 and 10. Bottom: Histogram of pixel intensities corresponding to the timepoints above. b) Decrease in fluorescence can be observed after 10 minutes when imaging at more than 50 mW (blue trace); signals are stable at lower intensities (black, cyan traces).*

### Figure S2 – Calibration steps

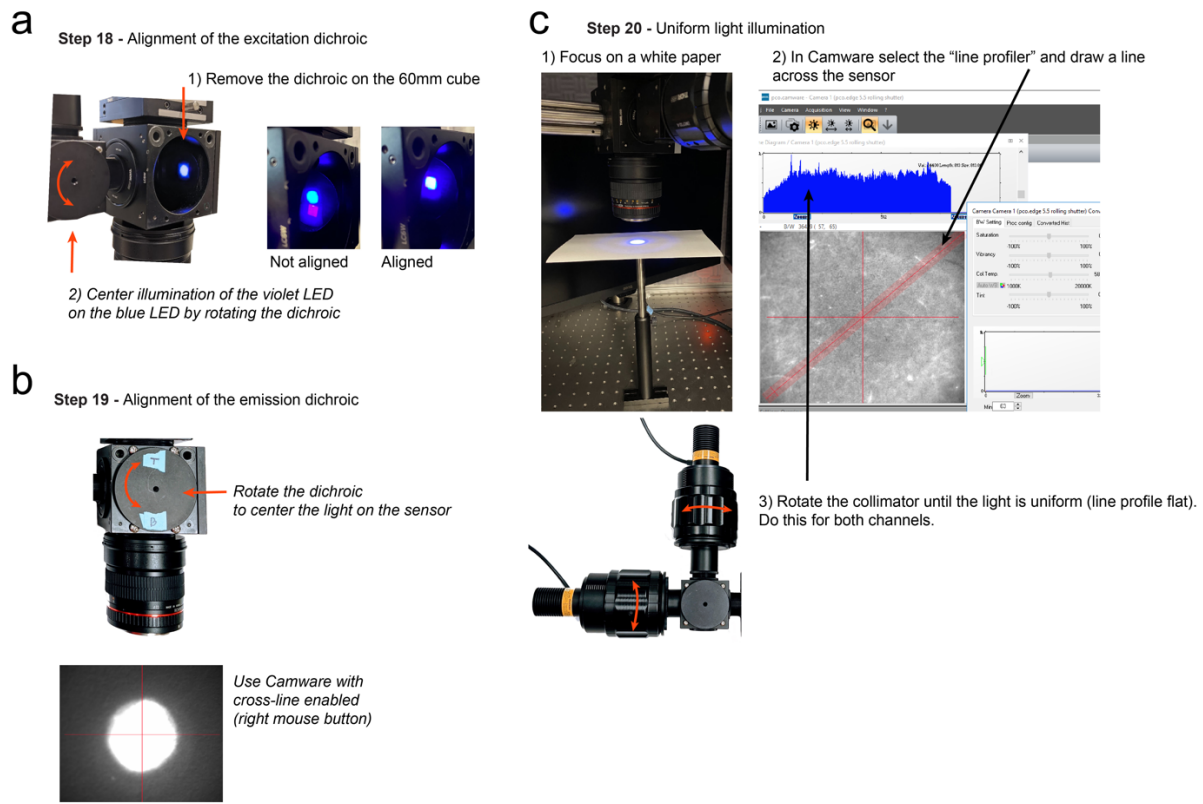

*Figure S2 – **Steps for setup calibration** a) Alignment of the excitation dichroic. b) Alignment of the emission dichroic using camware c) Procedure for obtaining uniform illumination with the camware software and the line profiler.*